\documentstyle[prd,aps,floats,times,epsfig,epsf]{revtex}
\draft
\input epsf
\def\be{\begin{equation}}
\def\ee{\end{equation}}
\def\beqa{\begin{eqnarray}}
\def\eeqa{\end{eqnarray}}
\def\etal{et. al.~}

\def\PRD{Phys. Rev. D~}
\def\ASTR{Astrophys. J.~}
\def\PLB{Phys. Lett. B~}

\def\PRL{Phys. Rev. Lett.~}

\begin{document}
\title{Dissipative Fluid in Brans Dicke theory and late time acceleration}
\author{A. A. Sen$^{\dagger}$, S.Sen$^{*}$ and S.Sethi$^{\ddagger}$}
\address{Harish-Chandra Research Institute, Chhatnag Road, Jhusi. Allahabad 211 019 
India}
\maketitle

\begin{abstract}
We have investigated the possibility of having a late time accelerated expansion phase for the universe. We have used a dissipative fluid in Brans-Dicke(BD) theory for this purpose. The model does not involve any potential for the BD scalar field. We have obtained the best fit values for the different parameters in our model by comparing our model predictions with SNIa data and the also with the data from the ultra-compact radio sources.
\end{abstract}

\pacs{PACS Number(s): 04.20Jb, 98.80Hw \hfill MRI-P-001001}
\vskip 2pc
A number of recent observations \cite{flat} suggest that the $\Omega_m$, the ratio of the matter density(baryonic+dark) to the critical density, is significantly less than unity suggesting that either the universe is open or that there is some other sources of this missing energy which makes $\Omega_{total}\sim 1$. The recent findings of BOOMERANG experiments \cite{BOOM} strongly suggests the second possibility of a flat universe. At the same time, the measurements of the luminosity-redshift relations observed for the 50 newly discovered type Ia supernova with redshift $z>0.35$ \cite{super} predicts that at present the universe is expanding in an accelerated fashion suggesting the existence of the total negative pressure for the universe.

One of the possibilities is the $\Lambda$CDM model consisting a mixture of vacuum energy or cosmological constant $\Lambda$ and cold dark matter(CDM). But as the vacuum energy is constant and the matter energy density decreases, it is necessary that their ratio must be set to a specific infinitesimally small value ($10^{-120})$ in the early universe so as to nearly coincide today. This is so called ``{\it cosmic coincidence}'' problem. Another possibility is ``{\it Quintessence}'' \cite{quint}, a dynamical, slowly evolving, spatially inhomogeneous component of energy density with negative pressure. An example is a time dependent scalar field slowly rolling down its potential \cite{scal}. Recently a new form of the quintessence called ``{\it tracker field}'' has been proposed to solve the cosmic coincidence problem. It has an equation of motion with an attractor like solution in a sense that for a wide range of initial conditions the equation of motion converges to the same solution \cite{track}.
There are a number of quintessence models which have been put forward and most of which involve a minimally coupled scalar field with potentials either exponential \cite{exp} or power law one \cite{power}. There are also treatments with the nonminimally coupled scalar fields with different type of potentials \cite{nonm}. It has been shown by Di Pietro and Demaret \cite{pietro} that for constant scalar field equation of state, which is a good approximation for a {\it tracker field} solution, the field equations and the conservation equations strongly constrain the scalar potential and the widely used potentials for quintessence such as the inverse power law one, exponential, and the cosine form, are incompatible with these constraints. Hence  it may be worthwhile to search for a model which will not involve any potential arising from a particle physics scale.

Negative pressure can also occur if the CDM fluid is not a perfect fluid but a dissipative one.
Recently it has been proposed that the CDM must  self interact in order to explain the detailed structure of the galactic halos \cite{CDM}. This self interaction will naturally create a viscous pressure whose magnitude will depend on the mean free path of the CDM particles. In a very recent work Chimento et.al have shown that a mixture of minimally coupled self interacting scalar field and a perfect fluid is unable to drive the accelerated expansion and solve the cosmic coincidence problem at the same time \cite{chimento},
while a mixture of a dissipative CDM with bulk viscosity and a minimally coupled self interacting scalar field can successfully drive an accelerated expansion and can solve the cosmic coincidence problem simultaneously. An effective negative pressure in CDM can also be created from Cosmic-anti friction which is closely related  with  the particle production out of gravitational field and one can have a similar dynamics like $\Lambda$CDM model as a special case of this Cosmic-antifriction \cite{zim}.

The present work investigates the possibility of obtaining an accelerated universe in Brans-Dicke (BD) theory with a dissipative fluid. Previously Bartolami \etal and Sen \etal \cite{nonm} have investigated such possibility in BD theory with a perfect fluid. But both of them have considered the potential for the BD scalar field itself which was not so in the original BD theory \cite{BD}. But in this work, we have not considered any potential for the BD scalar field. We have compared our solutions with the experimental data \cite{super} to constrain the different parameters in our model.  This simple enough model can be useful if one has to explain the quintessence model without scalar field potential.


For a flat FRW universe, with a scale factor $R(t)$, assuming the matter content is a dissipative fluid with only bulk viscosity, the BD field equations are:
\be
3{\dot R^2\over{R^2}} = {\rho_{m}\over{\phi}}+{\rho_{\phi}\over{\phi}}
\ee
\be
2{\ddot{R}\over{R}}+{\dot R^2\over{R^2}} = -{(p_{m}+\pi)\over{\phi}}-{p_{\phi}\over{\phi}}
\ee
\be
{\ddot{\phi}}+3{\dot
R\over{R}}{\dot\phi}={{\rho_{m}-3p_{m}-3\pi}\over{2\omega+3}}
\ee
where $\rho_{\phi}$ and $p_{\phi}$ are the energy density and pressure corresponding to the BD scalar field and is given by
\be
\rho_\phi=
\left[{\omega\over{2}}{\dot\phi^2\over{\phi}} -3{\dot{R}\over{R}}\dot{\phi}\right]
\ee
\be
p_\phi=\left[{\omega\over{2}}{\dot\phi^2\over{\phi}}+\ddot\phi
+2{\dot R\over{R}}\dot\phi\right]
\ee
The energy conservation equation for the matter field, which is not an independent equation but can be obtained using (1)-(3) is given by
\be
\dot\rho+3{\dot R\over{R}}(\rho_{m}+p_{m}+\pi)=0
\ee
We are considering a late time matter dominated universe hence $p_{m}=0$ in our case. 

Dissipative effects in FRW cosmology i.e negative $\pi$ can be modelled in two ways:
Firstly the conventional bulk viscous effect in a FRW universe can be modelled within the framework of nonequilibrium thermodynamics proposed by Israel and Stewart \cite{IS}. In this theory, the transport equation for the bulk viscous pressure $\pi$ takes the form
\be
\pi + \tau\dot\pi = - 3\zeta H - \frac{\tau\pi}{2}
\left[ 3H + \frac{\dot\tau}{\tau} - \frac{\dot T}{T} - 
\frac{\dot\zeta}{\zeta}\right]
\ee
where the positive definite quantity $\zeta$ stands for the coefficient for the bulk viscosity, $T$ is the temperature of the fluid, and $\tau$ is the relaxation time associated with the dissipative effect i.e. the time taken by the system to reach the equilibrium state if the dissipative effect is suddenly switched off. Provided the factor in the square bracket is small, one can approximate the equation (7) as a simple form 
\be
\pi + \tau\dot\pi = - 3\zeta H
\ee
which is widely used in the literature.  One can  also assume that the viscous effects are not so large as observations seems to rule out huge entropy productions in large scales \cite{entr}. The relation $\tau={\zeta\over{\rho_{m}}}$ can be assumed so as to ensure the viscous signal does not exceed the speed of light \cite{zeta} and also $(\tau H)^{-1}=\nu$ where $\nu>1$ for a consistent hydrodynamical description  for the fluid \cite{tau}. With these assumptions, equation (8) becomes 
\be
\nu H + {\dot{\pi}\over{\pi}} = - {3\rho_{m}H\over{\pi}}
\ee

Also as demonstrated in a recent paper by Zimdahl \etal ~\cite{zim} one can also have a negative $\pi$ if there exists a particle number non conserving interaction inside the matter. This may be due to the particle production out of gravitational field. In this case, the CDM is not a conventional disspative fluid, but a perfect fluid with varying particle number. Substantial particle production is an event that occurs in the early universe. But Zimdahl \etal have shown that extremely small particle production rate can also cause the sufficiently negative $\pi$ to violate the strong energy condition.

In our case, we are not {\it apriori} assuming any specific model for this negative $\pi$, rather only assuming the existence of a negative $\pi$, we have investigated the possibility of having the accelerated phase of the universe in BD theory, which is comparable with the observational estimates.
Using (1)-(3) one can write
\be
6{\ddot{R}\over{R}}+6{\dot{R}^{2}\over{R}^{2}}=2\omega {\ddot{\phi}\over{\phi}}+
6\omega {\dot{R}\dot{\phi}\over{R\phi}}-\omega {\dot{\phi}^{2}\over{\phi^{2}}}
\ee
To solve the system of equations we have assumed the following relation between the scale factor $R(t)$ and the BD scalar field $\phi$:
\be
\phi = AR^{\alpha}
\ee
where $A$ and $\alpha$ are constants. With (11), equation (10) becomes
\be
\dot{H}+\beta H^{2} = 0
\ee
where $\beta = {12-\omega\alpha^{2}-6\omega\alpha\over{6-2\omega\alpha}}$. Equation (12) on integration yields
\be
R = R_{0}\left({t\over{t_{0}}}\right)^{1/\beta}
\ee 
 where $R_{0}$ and $t_{0}$ are positive constants. One can identify $t_{0}$ as the present epoch i.e the age of the universe. Now from (11) one write
\be
\phi = \phi_{0}\left({t\over{t_{0}}}\right)^{\alpha/\beta}
\ee
where $\phi_{0} = AR_{0}^{\alpha}$. We will assume $\phi_{0}=1$ without any loss of generality in our subsequent calculations. The solutions for other physical quantites now become
$$
\rho_{m} = {1\over{\beta^{2}t_{0}^{2}}}\left({t\over{t_{0}}}\right)^{\alpha/\beta-2}\left[3-{\omega\over{2}}\alpha^{2}+3\alpha\right]
\eqno{(15)}
$$
$$
\rho_{\phi} = {1\over{\beta^{2}t_{0}^{2}}}\left({t\over{t_{0}}}\right)^{\alpha/\beta-2}\left[{\omega\over{2}}\alpha^{2}-3\alpha\right]
\eqno{(16)}
$$
$$
p_{\phi} = {1\over{\beta^{2}t_{0}^{2}}}\left({t\over{t_{0}}}\right)^{\alpha/\beta-2}\left[({\omega\over{2}}+1)\alpha^{2}+2\alpha-\alpha\beta\right]
\eqno{(17)}
$$
$$
\pi = -{1\over{\beta^{2}t_{0}^{2}}}\left({t\over{t_{0}}}\right)^{\alpha/\beta-2}\left[3-2\beta+({\omega\over{2}}+1)\alpha^{2}+2\alpha-\alpha\beta\right]
\eqno{(18)}
$$

\vskip 2pc
In these solutions we have essentially three parameters $\omega, \beta$ and $\alpha$, which are related (see just after equation 12). One has to ensure that the universe is accelerating i.e $0<\beta<1$ and also the density parameters for the matter and the scalar field are of the same order at present time i.e $\Omega_{0m}\sim\Omega_{0\phi}$. These will constrain the different parameters.

We obtain the best-fit value of $\beta$ by comparing our model
predictions with the SNIa data. We use the high-z data of the
Supernova Cosmology Project (SCP; Perlmutter {\it et
  al.} (1998))\cite{super} and the low-z data from Calan-Tololo survey (Hamuy {\it
  et al.} 1996)\cite{calan} for our study. Of the 60 data points, we use a
54 data points for our analysis (Fit C--D  of the SCP data;
for details of the excluded data points see Perlmutter {\it et al.}
1998)\cite{super}.
\begin{figure}[ht]
\centerline{
\epsfxsize=6cm\epsfysize=5cm\epsfbox{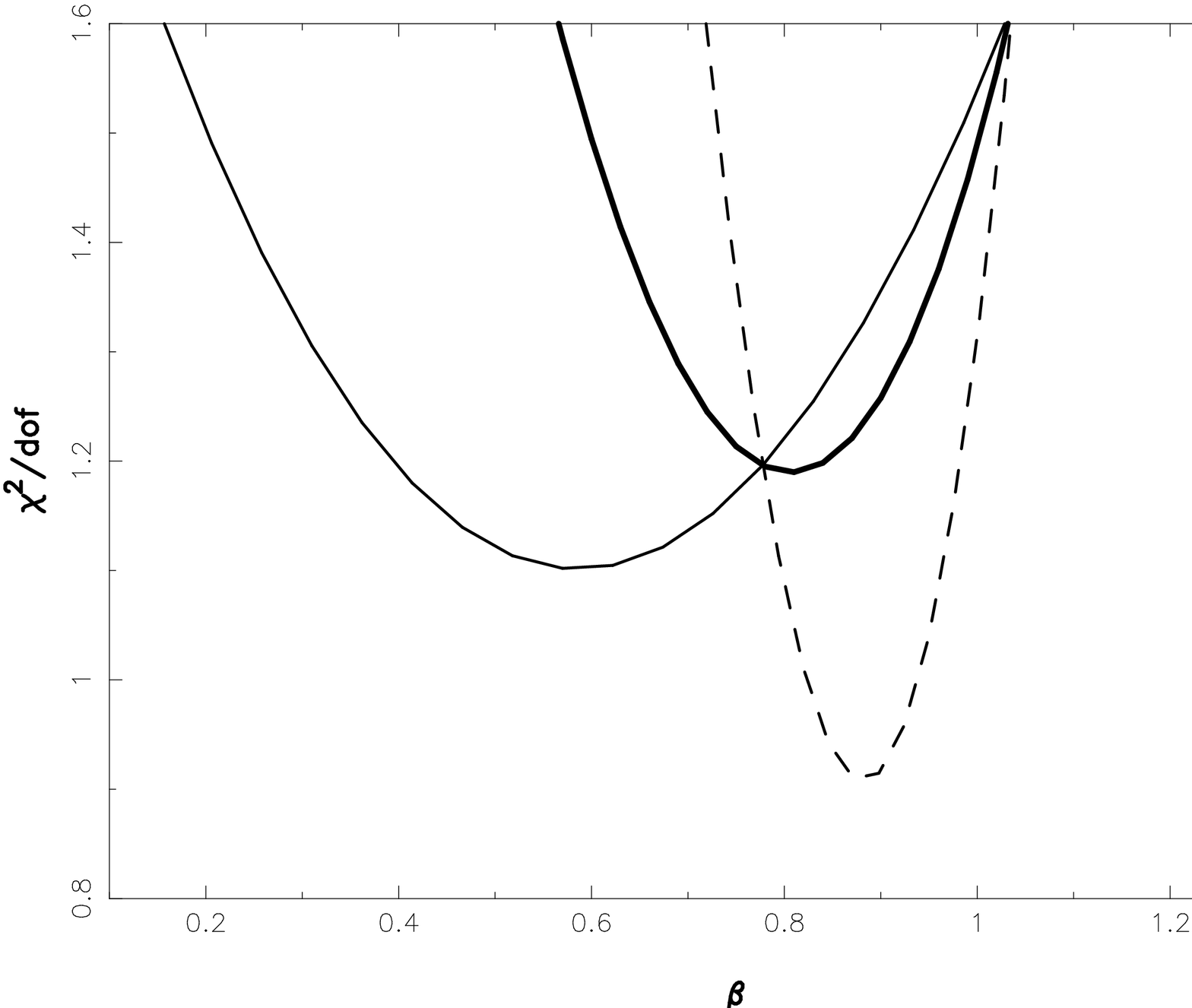}}
\vskip 1pc
\caption{$\chi^2/{\it dof}$ is shown for SNIa ({\it thin solid }
line) and Ultra-compact radio sources ({\it dahsed} line). The
{\it thick} solid line shows the result of joint analysis of the two
data sets}.
\label{Figure 1}
\end{figure}  
The SNIa data exist up to $z \simeq 0.9$. At larger redshifts,
we use the data of ultra-compact radio sources
($0.55 \le z \le 3.32$; 16 measurements) to constain the the value of
$\beta$\cite{Kel}.

The $\chi^2/{\it dof}$ of the comparison of our model with the two
data sets is shown in Figure~1. The joint analysis of the two
data set gives a best fit value $\beta = 0.8$ with $\chi^2/{\it dof} =
1.18$. The good-of-fit probability for the fit $Q = 0.12$, and
the 1$\sigma$ error on $\beta$ is $\Delta \beta = 0.05$.

In a very recent paper, using the data for the angular power spectrum of the cosmic microwave background obtained by MAXIMA-1 together with the measurements of high redshift Supernova, Balbi et.al \cite{balbi} have constrained density parameter for matter to be $0.25<\Omega_{m0}<0.5$. We have used this range of $\Omega_{m0}$ together with the value of $\beta$ obtained above by fitting our model with different observation, to constrain $\omega$ and $\alpha$.

We have plotted in figure~2  the parameters $\omega$ and $\alpha$ for $\beta=0.5$ and $\beta=0.8$ and also for $\Omega_{m0}=0.3$ and $\Omega_{m0}=0.5$. For this we have used the relation between $\beta$, $\omega$ and $\alpha$ and also equation (15). The two values of $\beta$ correspond to the current age of the universe  28 Gyr and 18 Gyr respectively where we have assumed that the present Hubble constant is $H_{0}\sim0.67\times 10^{-10}$ per year. One can see the ranges of the two parameters $\alpha$ and $\omega$, for which these values of $\beta$ and $\Omega_{m0}$ are consistent, are $-1.2\leq\alpha\leq -0.8$ and $-2.5\leq\omega\leq -1.5$ approximately. For these ranges of $\alpha$, the present day variation of $G$, $|{\dot{G}\over{G}}|_{0}=|\alpha| H_{0}<10^{-10}$ per year \cite{Solar}. One can also check from equations (17) and (18) using these range of parameters that both $p_{\phi}$ and $\pi$ remain negative in these ranges. 
Also one can also write from equation (9) 
$$
\nu={3[-5-5\alpha+\alpha^{2}(1+3\omega+\omega^{2})]\over{(\omega\alpha-3)(\alpha\omega+\alpha-1)}}
$$

In Order to have $\nu>1$ which is essential for hydrodynamical description if the CDM is assumed to be a conventional viscous fluid, one can not have a particular range for $\alpha$ and $\omega$ consistent with the ranges given above. Instead, for a particular value for $\alpha$ within the range given above, one can have a range for $\omega$. As an example, for $\alpha=-1.2$, the range for $\omega$ to have $\nu>1$ is $-2.25<\omega<-1.8$. For $\alpha=-1$,  it is $-2.5<\omega<-2$ and for $\alpha=-0.8$, it is $-2.5<\omega<-2.25$. One can see that these ranges of $\omega$ is consistent with the range shown in figure 2. 
\begin{figure}[ht]
\centerline{
\epsfxsize=7cm \epsfysize=5cm\epsfbox{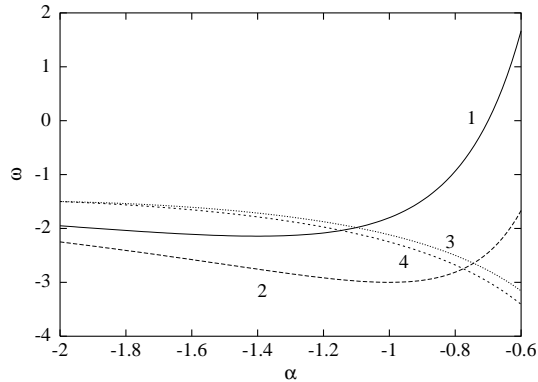}}
\caption{The parameter space $\omega$-$\alpha$ for different values of $\beta$ and $\Omega_{m0}$;
 1) $\Omega_{m0}=0.5$, 2) $\Omega_{m0}=0.3$, 3) $\beta=0.8$, 4) $\beta=0.5$}
\label{Figure 2}
\end{figure}

In conclusion, in recent years it has been shown that a mixture of perfect fluid and quintessence may be an
interesting candidate to explain a spatially flat universe currently
expanding in an accelerated manner. In these models, a minimally coupled
scalar field rolling down its potential has been used to drive the
accelerated expansion and also to account for the missing energy of
the universe. But all these models necessarily require several fine tuning
\cite{fine} of different parameters. Also in a very recent work,
Chimento et.al \cite{chimento} have shown that one can not simultaneously
solve the cosmic coincidence problem and have a late time acceleration in
FRW cosmology with a mixture of perfect fluid CDM and a Q-matter. On the
other hand, recent investigations have predicted that CDM should be self
interacting rather than collisionless, in order to successfully explain the
less dense galactic halos. Hence it is not unreasonable to think that this
self interaction may give rise to dissipative pressure $\pi$ at cosmological
scales. In their work, Chimento et.al have shown that a mixture of
dissipative CDM and a Q-matter can indeed explain the late time acceleration
and can solve the cosmic coincidence problem simultaneously. But all of these
quintessence models whether mixed with perfect fluid or dissipative fluid
suffer the problem of unwanted long range forces and the quintessence can
not be as homogeneous as it should \cite{problem}.

In this work, we have investigated the possibility of having a late time
 acceleration without any quintessence fields. We have used a dissipative
 CDM model in BD theory for this purpose. The viscous pressure together
 with negative pressure due to the BD scalar field  drive the late time
 acceleration and BD scalar field has been used to account for the missing
 energy of the universe.  We also have not used any potential for the BD
 scalar field unlike the other nonminimally coupled scalar field models in
 literatures \cite{nonm}. The model is simple enough and does not require
 much fine tuning. We have three arbitary parameters in our model which are
 related through an equation. We have constrained one of the parameters
 $\beta$ by fitting our model with the experimental data from Supernova and
 also from the ultra compact radio sources. The other two parameters $\alpha$
 and $\omega$ have been constrained using this value of $\beta$ and also
 assuming that the density parameters due to matter and the BD scalar field
 are of the same order today. These constraints give negative values for
 $\omega$.
However, standard  limit on  $\omega$ is $\omega>500$ to account the solar
system tests which sets tight constraints on Post Newtonian deviations from
general relativity\cite{Solar}:
$$
|\gamma_0-1|<2\times 10^{-3}~~~{\rm and}~~~|\beta_0-1|<2\times10^{-3}
$$
where $\beta_0$ and $\gamma_0$ denote the usual post Newtonian parameters.
But these constraints come from the weak field limit of the theory. 
One should also keep in mind that in extended
inflation, the model of La and Steinhardt\cite{la} worked provided that
$\omega$ takes a value close to 20 which is also not compatible with the
solar system tests. It should also be mentioned that in order to explain the
structure formation successfully in this scalar tensor theory the constraint
on $\omega$ is not at all compatible with the solar system tests\cite{struc}.
 A negative $\omega$ is also predicted  by the effective models comming from
 the Kaluza-Klein and superstring theories\cite{kk}. Hence it remains always a
problem to find a compatibility  between the astronomical observations and
cosmological requirements. 

The problem is to apply a theory in different scales (Astronomical and
cosmological) whereas experiments has been so far made only for
astronomical scales. We have applied the theory to cosmological scales
where still now there is no experimental tests for these scalar tensor
theories and future datas from supernova at higher redshift may confirm or
rule out existence of the scalar partner for the graviton.

It is also important to note that as $\beta$ is constant in our calculation, universe is always in the accelerating phase which seriously contradicts with the primeval nucleosynthesis and the structure formation scenario. One way to avoid this problem is to consider $\omega$ not a constant but a function of the scalar field $\phi$. In a recent interesting paper \cite{bpavon}, Banerjee and Pavon have shown that with $\omega$ as a polynomial function of $\phi$, one can get both radiation dominated era in the early time and accelerating phase in late time. But in that case also asymptotically $\omega$ acquires a small negative value 
to have a late time accelerating phase. 

Allowing $\omega$ to be a function of
$\phi$ or redshift $z$ to have both decelerating and accelerating phases at different times while local inhomogeneities giving rise to large value of $\omega$ consistent with solar system test, should  be the complete investigation.
 But this will
involve a detail computational efforts which is beyond the scope of this
paper. What we want to stress that if we can explain the quintessence in
Brans-Dicke theory without any potential( future observations will predict whether we
can or can not) then even if $\omega$ is scale dependent but then also in
some scale it has to satisfy the constraints given in our paper in order
to explain the late time acceleration and also the cosmic coincidence.


\end{document}